\documentclass[reprint,aps,amsmath,amssymb,prb]{revtex4-2}

\usepackage[version=4]{mhchem}
\usepackage{txfonts}
\usepackage{tipx}
\usepackage{threeparttable}
\usepackage{graphicx}
\usepackage{dcolumn}
\usepackage{bm}
\usepackage{color}
\usepackage{float}
\usepackage{booktabs}
\usepackage{listings}
\usepackage[ruled,linesnumbered]{algorithm2e}
\usepackage{algpseudocode}

\begin{document}
	
\preprint{APS/123-QED}

\title{Determination of melting temperature of hexagonal ice using Lee-Yang phase transition theory}

\author{Ling Liu}
\author{Yihua Dong}
\author{Qijun Ye} 
\email{qjye@pku.edu.cn}
\author{Xin-Zheng Li}
\email{xzli@pku.edu.cn}
\affiliation{Interdisciplinary Institute of Light-Element Quantum Materials, Research Center for Light-Element Advanced Materials, and Collaborative Innovation Center of Quantum Matter, Peking University, Beijing 100871, P. R. China}
\affiliation{State Key Laboratory for Artificial Microstructure and Mesoscopic Physics, Frontier Science Center for Nano-optoelectronics and School of Physics, Peking University, Beijing 100871, P. R. China}
\affiliation{Peking University Yangtze Delta Institute of Optoelectronics, Nantong, Jiangsu 226010, P. R. China}
\date{\today}

\begin{abstract}
Lee-Yang phase transition theory is a milestone in statistical physics.
Its applications in realistic systems, however, had been substantially hindered by availability of practical schemes to calculate the Lee-Yang zeros.
In this manuscript, we extend the scheme we have designed earlier [Phys. Rev. E 109, 024118 (2024)] and report simulation results for the melting temperature ($T$) of ice Ih 
under ambient pressure.
The enhanced sampling technique is shown to be crucial for accessing Lee-Yang zeros accurately.
The real and imaginary parts of our Lee-Yang edges demonstrate linear scaling of sizes, which can lead to a melting $T$ of 248.15 K for the TIP4P/2005 potential in the thermodynamic limit.
This result is in close quantitative agreement with previous coexistence simulations, achieved with cheaper computational costs and without prior knowledge of the phase transition.
With these, we demonstrate the applicability of Lee-Yang phase transition theory in realistic molecular systems, and provide a feasible scheme for high-throughput calculations in determining the phase transition temperature.
\end{abstract}

\maketitle

\section{Introduction}\label{sec1}

Phase and phase transition rank among the most ubiquitous phenomena in nature~\cite{Anderson1972,Wu1987,Debenedetti2001,Fong2004,Zhang2006,Sachdev2008,Carrasquilla2017,Ye2018}.
Since the early days of thermodynamics, theoretical description of them has became a fundamental challenge in physics~\cite{Stanley1987}.
One successful model for descriptions of the phase transition is the van der Waals (vdW) model, proposed in 1873~\cite{van1873}.
With it, one can explain the transition of the same matter between gas and liquid phases, even before the nature of the interactions between molecules is revealed and
the word of ``phase'' was proposed~\cite{Gibbs1875,Gibbs1877,Pauling1960}.
Starting from the beginning of the last century, phases and phase transitions have been systematically studied in magnetic systems, especially on the Ising model~\cite{Lenz1920,Ising1925,Onsager1944}.
In 1952, based on theoretical analysis of the Ising model and the lattice gas model, C. N. Yang and T. D. Lee proposed a theory for descriptions of the 
phase transitions using the zeros of the partition function, i.e. the Lee-Yang zeros on the complex plane of the thermodynamic state functions~\cite{Yang1952,Lee1952}.
This provides a beautiful mathematical framework for theoretical description of the phase transitions~\cite{Fisher1965}.
Applications of this theory, however, have been largely restricted to simple model systems, due to the lack of numerical scheme for calculations of the Lee-Yang zeros in
realistic systems~\cite{Flindt2013,Flindt2019,Flindt2022,Lee2013,van2015,Gnatenko2017}.
Parallel to understanding phase transitions using models, theoretical descriptions have also benefited from developments of molecular simulation methods since the mid-20th century~\cite{Frenkel2002}.
The advent of computers, especially supercomputers, have enabled numerical schemes like the molecular dynamics (MD)~\cite{Alder1958,Battimelli2018}, Monte Carlo (MC)~\cite{Metropolis1953,Frenkel2004}, and first-principles electronic structures methods~\cite{Car1985,Tuckerman2005}, transforming material science by allowing researchers to predict material properties with unprecedented precision and efficiency. 
In the context of phase transition studies, free-energy calculations and coexistence simulations have been the most popular methods for determining the phase boundaries.
The former aims to find the thermodynamics conditions where the free energies of the competing phases are equal~\cite{Frenkel1985,Piaggi2020,Piaggi2021,Gartner2022,Yang2021}, while the latter 
relies on MD simulations that can maintain the coexistence of the competing phases, starting from a box where they are in contact~\cite{Alfe2003,Gallo2017,Vega2022,Zhu2021}.
Both methods have achieved significant success in investigations of the realistic systems, such as water-ice transitions~\cite{Piaggi2020,Piaggi2021,Gallo2017,Vega2022}, liquid-liquid phase transitions~\cite{Gartner2022,Yang2021} and melting of high-pressure hydrogen~\cite{Pickard2012,Chen2014}.
Limitations of them, however, are also obvious.
A prominent one is that they typically require multiple simulations under different $T$ and $p$ conditions, rendering the study of phase boundaries a trial-and-error process that can incur high computational costs.
In recent years, there has been a growing interest in combining molecular simulations with the calculation of Lee-Yang zeros for realistic systems.
The central challenge lies in accurately determining the density of states~(DOS), which is essential for calculations of the partition functions.
Rocha \textit{et al.} utilized the multicanonical MC method to obtain precise estimates of the DOS in the canonical ($NVT$) ensemble, enabling the determination of Lee-Yang zeros for finite polymer systems~\cite{Rocha2014,Rocha2017}.
Recently, our group designed a scheme to calculate Lee-Yang zeros from MD simulations of the isothermal–isobaric ($NpT$) ensemble~\cite{Ouyang2024}.
In this scheme, the DOS for enthalpy is derived from its probability distribution, which can be obtained from the $NpT$ trajectories.
We demonstrated that the projected Lee-Yang zeros in the supercritical region align well with the Widom lines (commonly interpreted as ``supercritical boundaries''), while those in the phase transition region coincide with the phase coexistence line~\cite{Ouyang2024}. 
This advance implies an alternative strategy to free-energy calculations and coexistence simulations for studying phase transitions in realistic condensed matter systems.
Significant numerical challenges, however, still remain, including the fact that high free-energy barriers in most phase transition processes may lead to insufficient sampling of the probability distribution. 
Addressing this issue requires integrating enhanced sampling methods into the framework.
In this manuscript, we report our new progresses in this direction, by employing the on-the-fly probability enhanced sampling (OPES) method to calculate the Lee-Yang zeros.
As an example, we apply this method to study the phase transition between liquid water and ice Ih, a process that has been well-studied using the free-energy methods and direct coexistence technique~\cite{Piaggi2020,Piaggi2021,Gallo2017,Vega2022}. 
We show that the melting $T$ of ice Ih can be determined from the Lee-Yang zeros calculated using a single OPES simulation, rather than conducting multiple simulations under varying thermodynamic conditions.
This approach significantly reduces the computational cost, requiring only post-processing of MD simulation results using Lee-Yang phase transition theory. The associated computational overhead is negligible, and a Python script is provided to perform the analysis.
%
%
We also demonstrate excellent numerical agreement in melting $T$ predictions, with results remaining robust against artificial parameters such as simulated temperatures and enthalpy bin size.
This implies that simulations can be conducted without knowing the range of transition temperature, provided sampling efficiency is properly maintained.
By extrapolating the Lee-Yang zeros to infinite system size, the real part of the Lee-Yang edge on the complex plane of $T$ reaches the melting $T$ in the thermodynamic limit, while the imaginary part diminishes almost to zero. 
In so doing, this work presents a numerical scheme for investigating phase transitions in condensed matter systems within a single shot and without prior knowledge of phase transition. 
The paper is organized as follows. 
In Sec.~\ref{sec2}, we give a brief introduction to the method for calculating Lee-Yang zeros using MD simulation results, introduce the collective variables used for 
the OPES simulations, and provide the computational details. 
In Sec.~\ref{sec3}, we present numerical results for determining the melting $T$ from Lee-Yang zeros, with the effects of simulated temperatures and finite system sizes discussed. 
The conclusions are given in Sec.~\ref{sec4}. 
For ease of implementation, a Python script for calculating Lee-Yang zeros is provided in the Appendix A. Appendix B discusses the geometrical correlations between the enthalpy distribution and the zeros.  

\section{Method and Computational Details}\label{sec2}

\subsection{Lee-Yang Phase Transition Theory}

\begin{figure}[htbp]
	\includegraphics[width=1.0\linewidth] {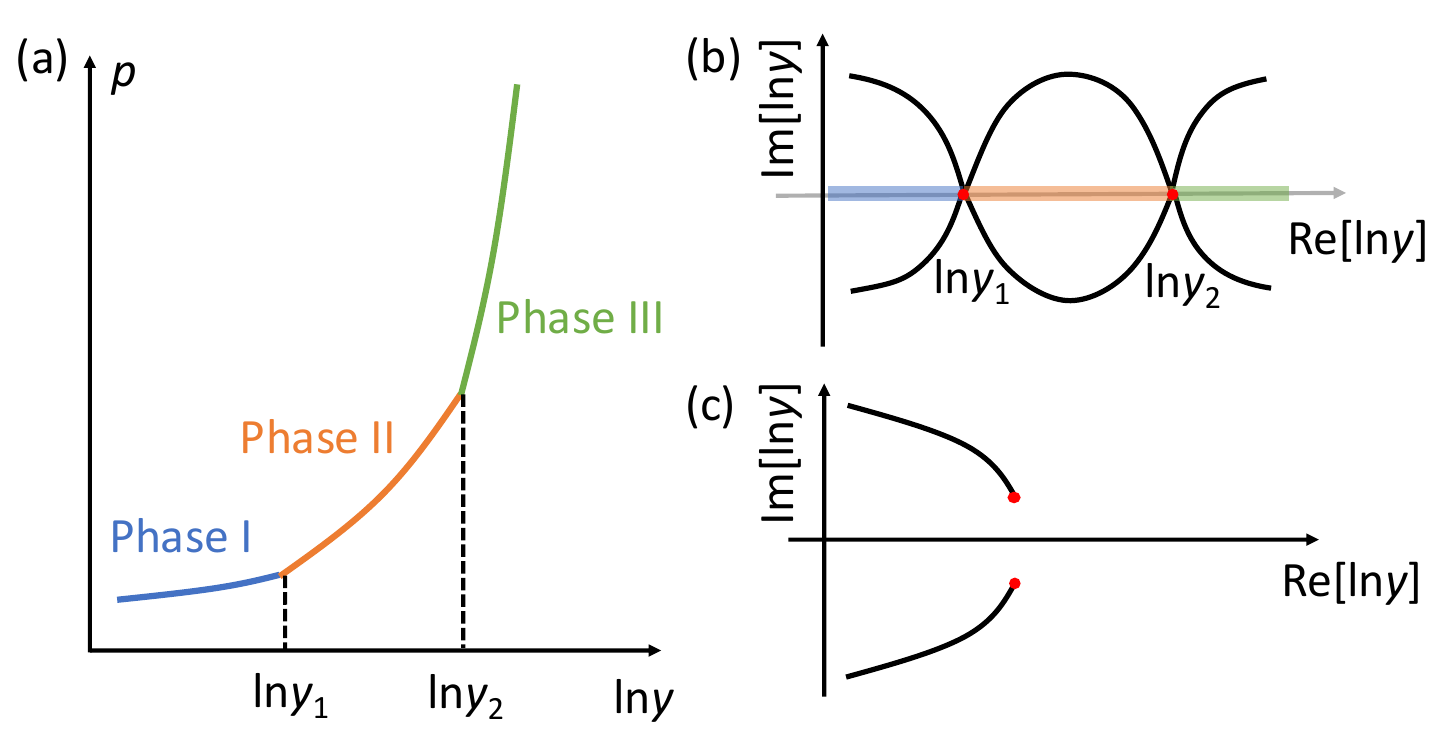}
	\caption{Schematic representation of the Lee-Yang phase transition theory. (a) Failure of the analytical property of the pressure $p$ as a function of $\text{ln}y$ in Eq.~(\ref{eq3}) in the case when phase transition happens at $\text{ln}y_1$ and $\text{ln}y_2$. (b) The distribution of the Lee-Yang zeros of $\text{ln}y$ in the complex plane of it when there are phase transitions, and (c) when there is no phase transition.}
	\label{fig:fig1}
\end{figure}

Phase transitions are characterized by the failure of an analytical function to describe a physical quantity using the state function, as abrupt changes observed in experiments~(Fig.~\ref{fig:fig1}~(a)).
A key factor in the mathematical description of this phenomenon, which can bring such a failure of the analytical function, is the ``singularity'' of the thermodynamic quantities.
In statistical physics, this is related to the ``zero'' of the ``partition function''~\cite{Yang1952}.
However, within the conventional understanding of the state functions, i.e. treating them as real numbers, the partition function is free of zero for all the physical thermodynamic conditions.
Following Ref.~[\onlinecite{Yang1952}], we take the lattice gas model as an example to illustrate this problem.
Consider a box with fixed volume $V$ and $T$, which can exchange atoms with a reservoir at a given chemical potential $\mu$. 
The
relative probability of having $N$ atoms in this box is given by
\begin{equation}\label{eq1}
	\frac{Q_N}{N!}  y^{N},
\end{equation}
where
\begin{equation}\label{eq2}
	Q_N=\int\cdots\int_V \text{d}r_1 \text{d}r_2\cdots \text{d}r_N \text{exp}[-U(r_1, r_2,\cdots, r_N)/(k_{\text{B}}T)],
\end{equation}
and 
\begin{equation}\label{eq3}
	y=(2\pi m k_{\text{B}} T/\hbar^2)^{3/2} \text{exp}[\mu/(k_{\text{B}} T)].
\end{equation}
This $Q_N$ is the configurational part of the partition function for $N$ atoms in this box, and $y$ is the thermal de Boroglie wavelength at $T$.
Then, the grand partition function writes,
\begin{equation}\label{eq4}
	Z_V=\sum_{N=0}^{M}\frac{Q_N}{N!}y^N,
\end{equation}
where $M$ represents the maximum number of atoms that can be contained in this box.
From Eqs.~(\ref{eq1}) to (\ref{eq4}), it is clear that with a physical chemical potential, i.e. a real $\mu$, the partition function $Z_V$ is always positive.
As such, a logical dilemma between conventional statistical physics, where zeros of the partition function are absent and consequently singularities of the
the thermodynamic quantities don't exist, and our experimental experiences with the phase transitions, where failure of analytical continuation often plays a central role, does exist.

The genius idea of the Lee-Yang phase transition theory, as proposed in Ref.~[\onlinecite{Yang1952}], is to treat the state functions, 
such as the $\text{ln}y$ in Fig.~\ref{fig:fig1}, as inherently complex rather than physically real ones.
%
%
As a result, the partition function, which remains positive for real $\ln y$s, emerges zeros for complex $\ln y$s.
%
Lee and Yang demonstrate that a phase transition can occur only in the thermodynamic limit, when some of these complex zeros approach a specific value on the positive real axis, marking the phase boundary (Fig.~\ref{fig:fig1}~(b)). 
These zeros, termed Lee-Yang zeros, correspond to complex chemical potential $\mu$ of the lattice gas model and magnetic field of the Ising model in the original papers of C. N. Yang and T. D. Lee~\cite{Yang1952, Lee1952}.
A few years later, this treatment of using complex number was extended to $T$, with the corresponding zeros called Fisher zeros~\cite{Fisher1965}.
We shall refer to these complex zeros collectively as Lee-Yang zeros.
For clarity, we distinguish them as $T$-zeros (Fisher zeros) or $\mu$-zeros (Lee-Yang zeros) to specify which state function is treated as complex.

Conventional theoretical studies of Lee-Yang phase transition theory have two obvious restrictions. 
The first one is that it was solely applied to studies of phase transition.
Last year, using van der Waals model, Ising model, and water system as examples, we demonstrated that this theory, based on Lee-Yang zeros, can also describe the supercritical matter~\cite{Ouyang2024}. 
The Lee-Yang edges, i.e. the Lee-Yang zeros closest to the real plane, locate at complex plane but largely correspond the Widom lines in the physical world~\cite{Ouyang2024}.
Given the significance of supercritical regions in the phase diagram, we believe this insight will motivate further investigations into supercritical states using the Lee-Yang phase transition theory.
The second one, as mentioned in the introduction, is that they were mostly based on model systems.
To address this, we developed a scheme to calculate the Lee-Yang zeros from MD simulations in our recent paper~\cite{Ouyang2024}.
However, calculating Lee-Yang zeros remains technically difficult, posing a significant obstacle to the broader applications and extensions of the Lee-Yang phase transition theory.
This challenge brings us to two key topics that will be explored in the next subsections, i.e. the scheme for calculating the Lee-Yang zeros from the MD simulations and how we improve it employing enhanced sampling techniques.

\subsection{Calculation of Lee-Yang Zeros Using MD}

We start from the scheme we have set up in Ref.~[\onlinecite{Ouyang2024}].
The central idea is that thermodynamic properties of a system should be described by complex phase diagrams, using the complex values of the state functions as its axes, rather than their real values.
For example, one should employ a 4-dimensional phase diagram denoted by $\tilde{T}$-$\tilde{p}$, where the tilde means complex numbers, 
rather than a 2-dimensional one denoted by $T$-$p$.
It has been shown in Ref.~[\onlinecite{Ouyang2024}] and later in this manuscript that this complex phase diagram enables an easier geometrical description of the controversial boundaries in conventional phase diagram.
Besides the $\tilde{T}$-$\tilde{p}$ phase diagram, there can be also $\tilde{T}$-$\tilde{h}$ and $\tilde{h}$-$\tilde{p}$ ones, describing the response of this system to other state functions such as the magnetic field $\tilde{h}$.
To reveal this 4-dimensional phase diagram in an intuitive manner, we choose to fix one state function at a real and physical value, and visualize the response of the system on the complex plane of another state function.
Taking the $\tilde{T}$-$\tilde{p}$ phase diagram as an example, we ran a series of $NpT$ MD simulations~\cite{Ouyang2024}.
To visualize the response of the system to $T$, we will fix $p$ and treat $T$ as a complex number $\tilde{T}$.
The partition function as a function of $T$ writes as an integral of the enthalpy $H$
\begin{equation}\label{eq5}
	Z= \int_H \rho(H)e^{-\beta H} \mathrm{d}H,
\end{equation}
where $\beta=1/(k_\text{B}T)$, and $\rho(H)$ is the DOS of $H$.
Then, the partition function in Eq.~\ref{eq5} is analytically continued as $Z_p(\tilde{T})$, replacing $T$ with $\tilde{T}$ while keeping all the other terms unchanged.
Similarly, to visualize the response of the system to $p$, we will fix $T$ and treat $p$ as a complex number $\tilde{p}$.
The partition function turns to be an integral of the volume $V$, as
\begin{equation}\label{eq6}
	Z= \int_V \rho(V)e^{-\beta pV} \mathrm{d}V,
\end{equation}
where $\rho(V)$ is the DOS of $V$, and $Z$ is analytically continued as $Z_T(\tilde{p})$ accordingly.
The partition function $Z$ can be reconstructed numerically, as $\rho(H)$ and $\rho(V)$ can be obtained from $NpT$ simulations.
Thus, using $Z_p(\tilde{T})$ and $Z_T(\tilde{p})$, one could analyze the structures of $Z$ and the Lee-Yang zeros.
As we focus on the melting $T$ in this work, we detail the structure of $Z_p(\tilde{T})$ and analyze the zeros of the partition function on the complex plane of $\tilde{T}$
in the following discussions, while the determination of the $\tilde{p}$-zeros follows a similar manner~\cite{Ouyang2024}.
In this case, $T$ becomes $\tilde{T}=T+i\tau$, with $\tau$ meaning the magnitude of its imaginary part.
The aim is to search for the roots of $Z_p(\tilde{T})=0$.
The integral in Eq.~(\ref{eq5}) will be discretized and represented as a polynomial.
Using a bin size $\Delta H$ and hence $H_k=H_0+k\Delta H$, the partition function can be rewritten as 

\begin{equation}\label{eq7}
	Z_p(\tilde{T})= \sum_k \rho_k e^{-\tilde{\beta} H_k } \Delta H =   \Delta H e^{-\tilde{\beta} H_0} \sum_k \rho_k [e^{-\tilde{\beta} \Delta H}]^k,
\end{equation}
The density of states $\rho(H)$ can be related to the probability distribution of enthalpy $g(H)$, by $\rho(H)=g(H)e^{\beta H}$. Then, we have
\begin{equation}\label{eq8}
	Z_p(\tilde{T}) =  \Delta H e^{-(\tilde{\beta}-\beta) H_0} \sum_k g_k  [e^{-(\tilde{\beta}-\beta) \Delta H}]^k.
\end{equation}
Note that $g_k = g(H_k)$ is the probability distribution of $H_k$ sampled at physical $T$ and $p$. 
After ignoring the constant and nonzero terms, the partition function in Eq.~(\ref{eq8}) is obviously proportional to a polynomial of $y = e^{-(\tilde{\beta}-\beta) \Delta H}$, in form of  
\begin{equation}
 Z_p(\tilde{T})  \sim \sum_k g_k  y^k \sim  \prod_n (y-y^*_n),
	\label{eq:poly}
\end{equation}
where $g_k$ serves as the polynomial coefficient.
In so doing, the roots of polynomial $y^*_n$ can be solved numerically with a Python script, provided in the Appendix A. 

\subsection{Selection of Collective Variables}
\begin{figure}[htbp]
	\includegraphics[width=0.75\linewidth] {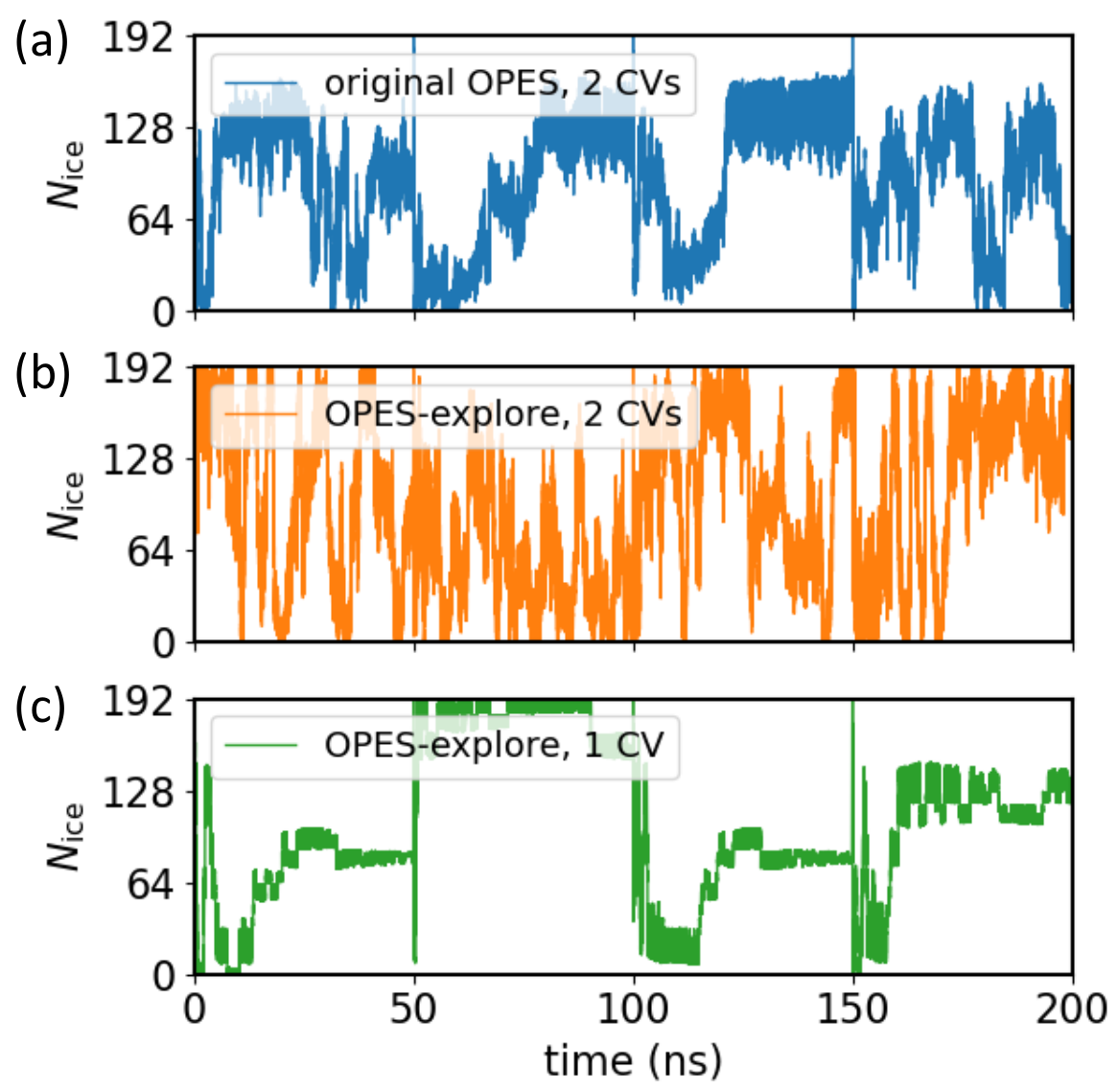}
	\caption{Evolution of $N_\text{ice}$ along with the simulation time for a 192-water system at 270 K using (a) the original OPES  and (b) OPES-explore, with bias applied to two CVs ($N_\text{ice}$ and $s_2$). (c) The result of the OPES-explore simulation using only one CV ($N_\text{ice}$) is shown for comparison. Note that the data from four multiple walkers are concatenated and presented in a single picture. }
	\label{fig:cvcomp}
\end{figure}

The above analysis is reliable as long as $g(H)$ converges in the $NpT$ simulations at the specified $(T,p)$.
However, this requirement has posed a serious challenge in practical MD simulations, especially for systems with complex liquid structures~\cite{Palmer2014,Piaggi2020,Gartner2022}.
To this end, we employ enhanced sampling techniques to obtain the enthalpy probability distributions. 
Using the rigid TIP4P/2005 water model~\cite{Vega2005}, we observe that the original OPES method~\cite{Invernizzi2020}, though efficient in achieving 
rapid convergence~\cite{Invernizzi2022}, fails to drive the system to a fully crystalline state, with only about 80\% of the water molecules freezing (Fig.~\ref{fig:cvcomp}~(a)). 
In contrast, a variant of the OPES method, i.e. the OPES-explore approach~\cite{Invernizzi2022}, performs significantly better, successfully achieving complete freezing (Fig.~\ref{fig:cvcomp}~(b)).
This is the first technical detail we want to emphasize.
Then, like all the adaptive-bias methods, appropriate collective variables (CVs) should be introduced in OPES-explore to distinguish between the liquid water and ice Ih, 
thereby facilitating both the crystallization and melting of ice phase. 
Here, we use the Environment Similarity CV that reflects the number of ice-like molecules in the simulation box~\cite{Piaggi2020,Piaggi2021}, ranging from 0 to $N$, the total number of water molecules. 
The construction of this CV involves \textit{three steps}.
\textit{First}, a similarity metric function $k_{\chi_l}(\chi)$ is defined to quantify the similarity between an environment $\chi$ around a given water and a reference environment $\chi_l$, where 
\begin{equation}
  k_{\chi_l}(\chi)=\frac{1}{n} \sum_{i\in \chi_l} \sum_{j \in \chi} \exp\left(  - \frac{ |\textbf{r}_j-\textbf{r}_i^0|^2}{4\sigma^2} \right).
  	  \label{eq:kchi}
\end{equation}
Here, $\textbf{r}_i^0$ and $\textbf{r}_j$ are the positions of neighboring atoms relative to the central oxygen atom in environments $\chi_l$ and $\chi$, respectively, $\sigma$ is a 
broadening parameter of Gaussian, and $n$ is the total number of neighbors used for normalization. 
The configurations of ice Ih are generated using GenIce2~\cite{Matsumoto2018,Matsumoto2024}. 
By inspecting these ice configurations with the Environment Finder tool~\cite{Piaggi2019,Piaggi2021b}, we identify four reference environments, each including 17 nearest 
neighbors up to the second solvation shell, with equivalent structures but different orientations. 
The \textit{second} step is to find out the best match with these four reference environments.
A smooth maximum function is used to ensure the CVs are continuous and differentiable, as required in driving OPES-explore simulations, by
\begin{equation}
	\begin{aligned}
 k(\chi) &= \max \{k_{\chi_1}(\chi),k_{\chi_2}(\chi),k_{\chi_3}(\chi),k_{\chi_4}(\chi)\}  \\
   &\approx \frac{1}{\lambda} \log \left(\sum_{l=1}^4  \exp(\lambda k_{\chi_l}(\chi))\right).
   \end{aligned}
\end{equation}
Here, $\lambda$ is set to 100 to make the smooth approximation solid. The Gaussian broadening $\sigma$ is set to 0.052 \AA~to make a well separation of ice-like and non-ice-like molecules in the $k(\chi)$ distribution~\cite{Piaggi2020}.
\textit{Third}, the similarity CV ($N_\text{ice}$) is defined as the number of ice-like molecules which satisfy $k(\chi^i) > \kappa$, where $\chi^i$ is the environment of water $i$, and the threshold $\kappa$ is set to 0.5 based on the $k(\chi)$ distribution~\cite{Piaggi2020}. 
Using the formula of a switching function, $N_\text{ice}$ can be expressed as,
\begin{equation}
  N_\text{ice} =N - \sum_{i=1}^{N} \frac{1-(k(\chi^i)/\kappa)^{12} }{1-(k(\chi^i)/\kappa)^{24}}.
\end{equation}
One key feature of this CV is that it enforces a specific lattice orientation during the ice crystallization process. 
We note that it has achieved success in studying the phase transition between liquid water and ice under the TIP4P/Ice water model~\cite{Piaggi2020}, as well as in the \textit{ab initio} MD simulations using SCAN functional and machine learning potentials~\cite{Piaggi2021}. 
However, our preliminary simulations of TIP4P/2005 water system with 192 molecules show that, when driven solely by this CV, the transition frequencies between 
water and ice basins are too low to ensure sufficient sampling and the convergence of thermodynamic properties (Fig.~\ref{fig:cvcomp}~(c)). 
This indicates that the barrier between the solid and liquid phases is not sufficiently removed by the bias of $N_\text{ice}$. 
%
%

Therefore, we introduce another CV as the translational entropy $s_2$, which has proven powerful in driving the homogeneous nucleation processes~\cite{Piaggi2017,Piaggi2018,Niu2019,Gartner2022}. 
It is formulated as a function of the instantaneous radial distribution function $g(r)$ of oxygen-oxygen,
\begin{equation}
	s_2 = -2\pi\rho k_\text{B} \int_{0}^{r_\text{max}} [g(r)\ln g(r)-g(r)+1]r^2 dr,
\end{equation}
where $\rho$ is the density of the system, and $r_\text{max}$ is the upper limit of the radial distance integration, set to 5 \AA. 
The addition of CV $s_2$ leads to a noticeable increase in the transition frequencies, as shown in Fig.~\ref{fig:cvcomp}~(b). 
Thus, we use both $N_\text{ice}$ and $s_2$ to conduct all the OPES-explore simulations. 

\subsection{Computational Details}

All the OPES-explore simulations were carried out using the LAMMPS software~\cite{lammps2022} interfaced with PLUMED~\cite{plumed2014}. 
For each $NpT$ simulation, $T$ and $p$ were maintained using the N\'ose-Hoover thermostat and barostat~\cite{Nose1984,Hoover1985}, with the effective relaxation time of 0.1 ps and 1 ps. 
A timestep of 1 fs was used.
The cutoff for the Lennard-Jones and Coulomb interactions was set as 10 \AA, while the long-range Coulomb interactions were computed with the 
particle-particle particle-mesh (PPPM) solver~\cite{Hockney1988} in the reciprocal space at an accuracy of 10$^{-6}$.
Tail corrections to the energy and $p$ were added to include the long-range van der Waals interactions~\cite{Sun1998}. 
The bond lengths and angles of water were fixed using the SHAKE algorithm~\cite{Berendsen1977}.
To accelerate the convergence of sampling, the OPES-explore simulations were performed using four walkers~\cite{Parrinello2006}. 
Each walker was equilibrated for at least 50 ns. 
The bias potential was updated every 500 timesteps. 
The approximate free-energy barriers to be overcome for systems of 64, 128 and 192 molecules were set to 50, 100 and 150 kJ/mol, respectively. 
The bias factors for the target well-tempered distributions were automatically determined based on the predefined barrier heights~\cite{Invernizzi2022}. 
In addition to the primary CVs promoting the formation of ice Ih, we employ two extra CVs to prevent the emergence of undesirable configurations, i.e. structures misaligned with the prescribed lattice orientation and those corresponding to the cubic ice (Ic) phase, as detailed in Ref.~[\onlinecite{Piaggi2020}].
For OPES-explore simulations, we use a descriptor defined by the free-energy difference between liquid water and ice to evaluate the convergence of sampling. 
It is calculated for all the sampled configurations, as
\begin{equation}
	\Delta F= -\frac{1}{\beta} \ln \left(\frac{P_\text{ice}}{1-P_\text{ice}} \right)
\end{equation}
where $P_\text{ice}$ is the probability of configurations being in the ice phase. 
This $P_\text{ice}$ is computed by counting the number of configurations (after reweighting) with $N_\text{ice}$ larger than a threshold $N_{\text{th}}$, 
which distinguishes ice states from liquid water. 
Theoretically, $N_{\text{th}}$ should correspond to the dividing surface that separates the reactant and product regions. 
While in practice, the value of $\Delta F$ is largely insensitive to the choice of $N_{\text{th}}$, as configurations with high free-energy contribute minimally to $P_\text{ice}$. 
Thus, $N_{\text{th}}=N/2$ is used for simplicity~\cite{Piaggi2020}.

\begin{figure}[htbp]
	\includegraphics[width=9.04cm] {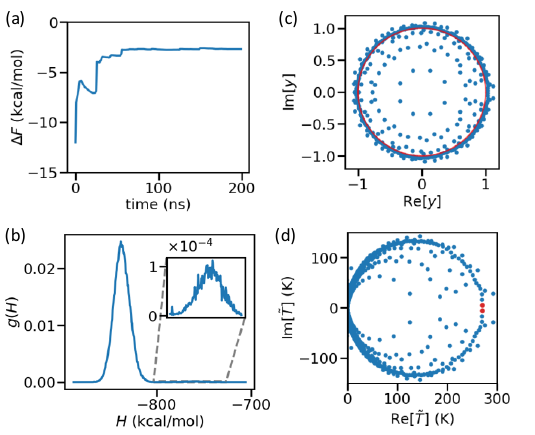}
	\caption{(a) The convergence of $\Delta F$ as a function of simulation time. (b) The probability distribution of $H$ reweighted from the total 200 ns trajectory of 4 walkers. $H$ is equal to $U+pV$, where $U$ and $V$ denotes the potential energy and volume, respectively. $p$ is fixed at 1 bar. The bin size $\Delta H$ is set to 0.5 kcal/mol. (c) The distribution of the roots of partition function expressed in polynomial form (Eq.~\ref{eq:poly}). The red line represents a unit circle centered at the origin. (d) Lee-Yang zeros in the complex $T$ plane. The red points highlighted refer to the zeros closest to the real axis. The distribution of zeros displays a symmetric pattern about the real axis. } 
	\label{fig:steps}
\end{figure}

\section{Results and Discussions}\label{sec3}

We start with the OPES-explore results of 64 water molecules at 260 K and 1 bar. 
In Fig.~\ref{fig:steps}(a), we show the convergence of $\Delta F$ for the trajectories of four walkers, each with a duration of 50 ns, as a function
of simulation time.
The curve reaches a plateau after 100~ns, indicating that the sampling is sufficient for the analysis of Lee-Yang zeros. 
The reweighted probability distribution of enthalpy, derived from the full trajectory, is presented in Fig.~\ref{fig:steps}~(b).
Double peaks, representing ice (lower $H$) and liquid (higher $H$), highlight a pronounced stability difference between the two phases, with the ice phase being notably predominant at 260 K. 
The roots $y^*_n$, referred to as $y$-zeros derived from Eq.~\ref{eq:poly}, lie primarily on a circle in the complex plane with a radius slightly greater than 1, 
as shown in Fig.~\ref{fig:steps}~(c). 
Due to the deviations of $H$ from a normal distribution caused by statistical errors, which is inherent in the numerical sampling techniques like 
MD or MC, some roots also lie outside or inside this circle.
Using the relation $y=e^{-(\tilde{\beta}-\beta) \Delta H}$, one can also convert $y$-zeros to $\tilde{T}$-zeros, i.e. the zeros on the complex plane of temperature, as
shown in Fig.~\ref{fig:steps}~(d). 
The $\tilde{T}$-zeros closest to the real axis, referred to as the $\tilde{T}$-edges (denoted by red points), are located at 269.15$\pm$5.57$i$~K. 
The real part of $\tilde{T}$-edges corresponds to the melting temperature, while the imaginary part reflects the ``offset'' from the thermodynamic limit, associated with the fact that a finite size of the simulation cell is used.

In Ref.~\onlinecite{Lee1952}, the Lee-Yang unit circle theorem was proposed for $y\sim e^{\mu/(k_\text{B}T)}$, which, according to their discussions, is applicable for a wide range of interatomic and intermolecular interactions.
Our numerical simulations of the roots $y=e^{-(\tilde{\beta}-\beta) \Delta H}$ for the phase transition between liquid water and ice confirm the circular distribution predicted by this theorem, as clearly shown in Fig.~\ref{fig:steps}~(c).
The thermodynamic state function, however, is $T$ instead of $y$. 
In Fig.~\ref{fig:steps}~(d), we can also see an approximate circular pattern for the $\tilde{T}$-zeros.
To understand this, we assume $y^*=r e^{i\theta}$.
Then it can be derived that the zeros of $\tilde{T}$ also lie on a circle, centered at $T_0/2$ with a radius $T_0/[2(1-k_\mathrm{B}T_0 \ln r/\Delta H)]$, where $T_0$ is the 
target temperature of the $NpT$ simulation.
Being on the right boundary of this circle, the real part of the $\tilde{T}$-edges can be approximately expressed as
\begin{equation}
	\text{Re}[\tilde{T}\text{-edges}] = \frac{T_0}{2} + \frac{T_0}{2(1-k_\mathrm{B}T_0 \ln r/\Delta H)}.
\end{equation}
However, one can still see subtle deviations from a circle shape in Fig.~\ref{fig:steps}~(d). 
A detailed discussion of the physicals meanings of these zeros features, analyzed through Gaussian distribution models, is provided in the Appendix B (with a figure in it).
Remarkably, this scheme allows determining the phase transition temperature even when the simulated $T_0$ is far  from the melting temperature.

\begin{figure}[htbp]
	\includegraphics[width=8.94cm] {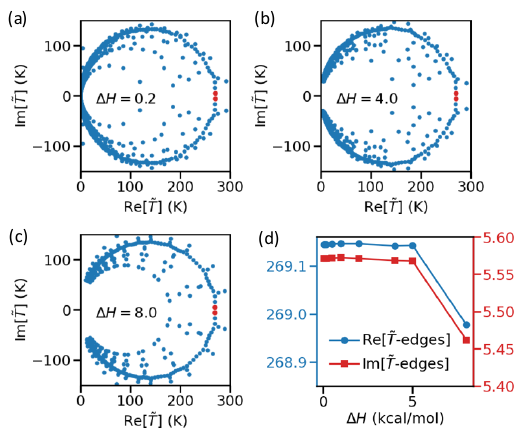}
	\caption{Distributions of $\tilde{T}$-zeros with bin sizes of (a) $\Delta H=0.2$ kcal/mol, (b) $\Delta H=4$ kcal/mol and (c) $\Delta H=8$ kcal/mol. (d) The real and imaginary parts of the $\tilde{T}$-edges as a function of $\Delta H$, represented by blue circles and red squares, respectively.  }
		\label{fig:dHtest}
\end{figure}

It is worth mentioning that the number of bins discretizing $H$ equals the number of $y$-zeros, according to Eq.~(\ref{eq:poly}). 
Consequently, a smaller bin size will lead to a denser distribution of zeros. 
To assess the influence of bin size on the distribution of Lee-Yang zeros, we vary $\Delta H$ across a range of values from 0.05 to 8 kcal/mol, with results shown in Fig.~\ref{fig:dHtest}. For $\Delta H \le 5$ kcal/mol, the $\tilde{T}$-zeros distributions exhibit good consistency (Fig.~\ref{fig:dHtest}~(a) and \ref{fig:dHtest}~(b)), and crucially, the variation in $\tilde{T}$-edges remains well within the statistical uncertainties (Fig.~\ref{fig:dHtest}~(d)), reinforcing the robustness of this approach. However, increasing $\Delta H$ to 8 kcal/mol results in a noticeable decrease in both the real and imaginary parts of the $\tilde{T}$-edges. Such large values of $\Delta H$ should be avoided, as they lead to overly coarse estimations of $g(H)$ and, consequently, the partition function. 

%
%

\begin{figure}[htbp]
	\includegraphics[width=9.0cm] {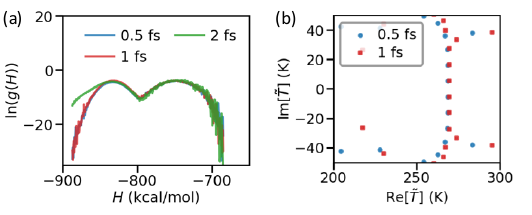}
	\caption{(a) Comparison of $H$ distributions simulated with different time steps for a 64-water system at 270 K. (b) A subset of $\tilde{T}$-zeros calculated from simulations with time step of 0.5 fs (blue circles) and 1 fs (red squares), with the $\tilde{T}$-edges located at 268.70$\pm$5.44$i$ K and 269.50$\pm$5.45$i$ K, respectively.    } 
	\label{fig:timestep}
\end{figure}

Special attention should be given to the choice of time step, $\Delta t$. 
When simulating with rigid water models, an common practice is to use a larger but more efficient time step as $\Delta t=2$ fs~\cite{Piaggi2020,Niu2019,Gallo2017,Vega2022}.
This is feasible since the high-frequency vibrations between bonded atoms are nearly frozen, such as by the SHAKE algorithm~\cite{Berendsen1977}. 
However, a recent study recommended a smaller time step ($\Delta t=0.5$ fs) to ensure proper equipartition in MD simulations involving rigid water models~\cite{Asthagiri2024}. 
We have conducted tests on the time step, with results shown in Fig.~\ref{fig:timestep}. 
Using a 2 fs time step leads to significant deviations from the Gaussian distribution for the enthalpy in ice phase, which might be related to the overestimation of the non-equilibrium structures during the sampling process.
In contrast, $H$ distribution obtained with a 1 fs time step closely matches the one with 0.5 fs (Fig.~\ref{fig:timestep}~(a)), resulting in consistent $\tilde{T}$-zeros, especially the $\tilde{T}$-edges (Fig.~\ref{fig:timestep}~(b)). 
Thus, our results suggest that $\Delta t=1$ fs is necessary for accurate simulations of this phase transition, at least when using the OPES-explore method. 

\begin{figure}[htbp]
	\includegraphics[width=9.0cm] {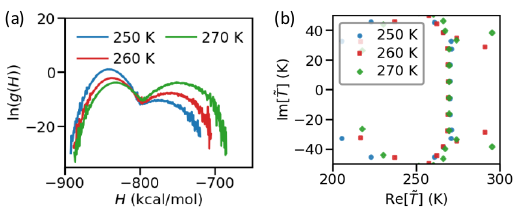}
	\caption{(a) The enthalpy probability distribution obtained from the OPES-explore simulations conducted at 250, 260 and 270 K. (b) A subset of $\tilde{T}$-zeros calculated from the simulations at 250 K (blue circles), 260 K (red squares) and 270 K (green diamonds).  } 
	\label{fig:temp}
\end{figure}

To ensure numerical stability of this scheme, we also performed OPES-explore simulations with target $T$s of 250 K and 270 K, in addition to 260 K, under the same $p$ (ambient pressure). 
%
%
The resulting probability distributions of $H$ at different temperatures are compared in Fig.~\ref{fig:temp}~(a). 
As $T$ decreases, the probability of configurations being in the liquid phase drops by several orders of magnitude, while the probability of ice phase significantly increases. 
Despite the distinct $H$ distributions, the $\tilde{T}$-zeros shown in Fig.~\ref{fig:temp}~(b) show highly consistent behaviors across these simulations.
Specifically, the $\tilde{T}$-edges obtained at 250, 260, 270 K are located at  270.22$\pm$5.71$i$, 269.15$\pm$5.57$i$ and 269.50$\pm$5.45$i$ K, respectively. 
This demonstrates that the melting temperature, determined from Lee-Yang zeros, is basically insensitive to the target $T$ of the $NpT$ simulation, as long as the probability distribution of $H$ is well-sampled. 
In general, one could determine the phase boundary within a single simulation by using OPES or other enhanced sampling methods, without prior knowledge of the rough temperature range of the phase transition.
This scheme enables an efficient exploration of the phase diagram without experimental inputs or artificial supervision, making it suitable for high-throughput calculations.

\begin{figure}[htbp]
	\includegraphics[width=9.0cm] {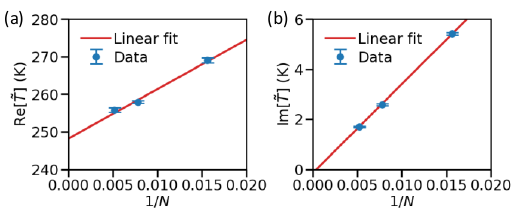}
	\caption{(a) The real part and (b) the imaginary part of $\tilde{T}$-edges as a function of inverse system size. The errors are calculated using 3 blocks, each with a trajectory length of 200 ns simulated at a temperature of 270 K. }
	\label{fig:size}
\end{figure}

To deduce the phase boundary in the thermodynamic limit, we further simulate larger systems containing 128 and 192 molecules. 
The calculated $\tilde{T}$-edges as a function of inverse system size are shown in Fig.~\ref{fig:size}. 
Both the real and imaginary components of $\tilde{T}$-edges exhibit linear relationships with $1/N$, in alignment with the finite size scaling predictions for first-order phase transitions~\cite{Binder1987}. 
Through linear fitting, the real part of the $T$ edges approaches a melting temperature of 248.15 K in the thermodynamic limit, while the imaginary part almost diminishes. 
The melting temperatures of ice Ih at 1 bar for TIP4P/2005 water model, obtained using different methods, are summarized in Table \ref{tb:Tm}. 
Our result is in quantitative agreement with those determined from the direct coexistence technique~\cite{Vega2022,Gallo2017}. 
The effectiveness of our method is built on the fact that Lee-Yang zeros is intrinsically equivalent to the partition function, as demonstrated by Eq.~\ref{eq:poly}.
Furthermore, its robustness for the insensitivity to parameters, and efficiency in requiring only a single simulation instead of a series of coexistence or heating/cooling simulations, should be emphasized. 
We anticipate further applications of this scheme in a wider range of realistic systems.

\begin{table}[htbp]
	\centering
	\caption{Comparison of the melting points of ice $\text{I}_h$ at ambient pressure for the TIP4P/2005 model, determined using different methods.  }
	\begin{tabular}{lcc}
		\toprule
		Method  & melting $T$ (K)  &  $N$ \\
		\midrule
		This work & 248.15 & $\infty$ \\
		Direct coexistence~\cite{Gallo2017} &  249.5 $\pm$ 0.1 & 16000 \\
		Direct coexistence~\cite{Vega2022} & 250 &  4000 \\
		\bottomrule
	\end{tabular}
\label{tb:Tm}
\end{table}

\section{Conclusions}\label{sec4}

We apply the Lee-Yang phase transition theory to determine the phase transition $T$ between ice Ih and liquid water. 
To approach ergodicity and obtain converged probability distributions, we conduct OPES-explore simulations using two CVs, i.e. the number of ice-like molecules $N_\text{ice}$ and the translational entropy $s_2$. 
By analyzing the probability distribution of enthalpy, we derive the $\tilde{T}$-zeros of the partition function. 
The zeros close to the real axis, especially the $\tilde{T}$-edges, are insensitive to the target $T$ of the $NpT$ simulation and the bin size $\Delta H$.
In the thermodynamic limit, we obtain a melting $T$ of 248.15 K, in good agreement with previous results.
By integrating the enhanced sampling method, this work establishes a practical and efficient scheme for determining phase transition temperature in realistic systems without prior knowledge of the phase transition.

\begin{acknowledgments}
	This work is supported by the National Science Foundation of China under Grants No. 12234001, No. 12204015, No. 12474215, and No. 62321004, the National Basic Research Program of China under Grants No. 2021YFA1400500 and No. 2022YFA1403500. We thank the supercomputer center at Peking University for computational resources. 
\end{acknowledgments}

\appendix


	


\section{Python script for calculating $\tilde{T}$-zeros}

\begin{algorithm}
	\renewcommand{\algorithmicrequire}{\textbf{Input:}}
	\renewcommand{\algorithmicensure}{\textbf{Output:}}
	\label{alg:polynomial_roots}
	\begin{algorithmic}[1]
		\Require File `histo' containing probability density data
		\Ensure Generate polynomial roots and temperature zeros
	
		\State \textbf{Initialize variables}
		\State $dH \gets 0.5$ \Comment{Bin size of energy (kcal/mol)}
		\State $T_0 \gets 270$ \Comment{Simulated temperature (K)}
		
		\State \textbf{Load probability density file}
		\State $H \gets \text{np.loadtxt(`histo')}$
		
		\State \textbf{Calculate roots of the polynomial}
		\State $roots \gets \text{np.polynomial.polynomial.polyroots}(H[:, 1])$
		\State $X \gets \text{Re}(roots)$
		\State $Y \gets \text{Im}(roots)$
		
		\State \textbf{Calculate temperature zeros}
		\State $ln\_values \gets \text{np.log}(roots)+2k\pi i$
		\State $Beta\_zeros \gets -ln\_values/dH + 1/(k_\text{B} \times T_0)$
		\State $T\_zeros \gets 1/(Beta\_zeros \times k_\text{B}  )$
		\State $Tx \gets \text{Re}(T\_zeros)$
		\State $Ty \gets \text{Im}(T\_zeros)$
		
	\end{algorithmic}
\end{algorithm}

\section{The geometrical correlations between the enthalpy distribution and zeros}

\begin{figure*}[htbp]
	\includegraphics[width=12.88cm] {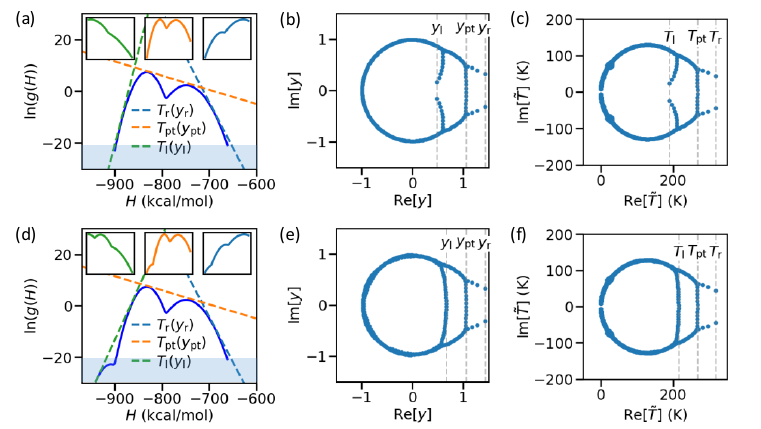}
	\caption{(a) A schematic plot of the probability distribution of enthalpy, modeled with two Gaussian functions representing the ice and liquid water phases, respectively. The cyan, orange and green dashed lines denotes the titled temperature axes. The insets display the rescaled $H$ distribution after reweighting to the target $T$ or $y$. The blue shadow depicts the area below the resolution threshold. (b) and (c) show the distribution of $y$-zeros and $\tilde{T}$-zeros. $T_\text{l}$($y_\text{l}$) and $T_\text{r}$($y_\text{r}$) are the real parts of the left and right branch, respectively, while $T_\text{pt}$($y_\text{pt}$) denotes the phase transition point. (d)(e)(f) correspond to (a)(b)(c), but include a subtle peak below the original sampling resolution. In (e) and (f), the left branch of zeros shift closer to the real axis compared to (b) and (c), signaling a new phase transition.}
	\label{fig:model}
\end{figure*}
In the practice of calculating zeros, one might observe more zeros approaching the real axis than the one relates to phase transition.
Here, we shall demonstrate that these additional zeros rise from limited sampling resolution and are thus unphysical.
In $NpT$ ensemble, the $H$ probability $g_\beta (H)$ observed at $T=\beta^{-1}$ can be scaled from a reference $g_{\beta_0}(H)$ at $T_0 =\beta_0^{-1}$ by a exponential  factor $e^{-(\beta - \beta_0)H}$.
To provide an intuitive understanding, we employ a geometrical explanation.
When the $g_\beta (H)$ is plotted on a logarithm scale, this factor manifests as a tilted temperature axis.
Since the choice of $T_0$ is arbitrary, we consider an example where $T_0$ is near the phase transition temperature, at which $g(H)$ exhibit a double peak lineshape, as shown by the blue curve in Fig.~\ref{fig:model}~(a).
Correspondingly, the calculated zeros form three branches, as shown by $y$-zeros in Fig.~\ref{fig:model}~(b) and $\tilde{T}$-zeros in Fig.~\ref{fig:model}~(c).
The central branch, denoted as $T_{\text{pt}}$($y_\text{pt}$), corresponds to the target phase transition, while the left and right one, termed by $T_\text{l}$($y_\text{l}$)  and $T_\text{r}$($y_\text{r}$), are unphysical.

In this geometrical framework, the original $T$-axis is horizontal, aligned with $\beta_0$, but it tilts to a slope corresponding to $\beta_i$ at $T_i$.
The interplay between the tilted $T$-axis and the $g(H)$ lineshape provides insights into the behavior of zeros.
When the tilted $T$-axis is tangent to both peaks, the slope exactly maps to the transition temperature $T_{\text{pt}}$, as shown as orange lines in Fig.~\ref{fig:model}~(a).
Below and above this $T_{\text{pt}}$, the tilted axis is tangent to either the left or right peak, respectively.
We represent the temperatures at the edges of the left and right branches of zeros, i.e., $T_\text{l}$ and $T_\text{r}$, with green and cyan dashed lines, respectively. 
$T_\text{l}$ and $T_\text{r}$ appear to be the endpoints beyond which the axis become intersect with the lineshape of $g(H)$  instead of being tangent to it.
The reweighted $g(H)$ for $T_\text{l}$, $T_{\text{pt}}$, and $T_\text{r}$ are plotted in the insets of Fig.~\ref{fig:model}~(a).
As the temperature increases, the lineshape switch between a single dominant peak and double peaks, reflecting that the system experiences a phase transition.
Thus, we connect the tangent behavior with the positions of edge zeros.

However, sampling techniques inherently overlook configurations with probabilities significantly below the resolution threshold.
In primitive sampling, the sample size $N$ determines that configurations with probability much less than $N^{-1}$ are rarely sampled.
While in enhanced sampling methods, configurations with $p \ll c\cdot N^{-1}$ are still out of reach, even though the efficiency has been enhanced by a factor of $c$.
This limitation is particularly pronounced on a logarithm scale.
In fact, additional peaks may exist below the resolution threshold and are undetected.
For example, as show in Fig.~\ref{fig:model}~(d), one would perceive a new peak when the sampling was enhanced by ten orders of magnitude.
Before the axis tilts to $T_\text{l}$, it become tangent to both the new peak and original left peak.
Correspondingly, we observe that the zeros shift closer to the real axis, indicating a real phase transition, while $T_{\text{pt}}$ and $T_\text{r}$ remain unchanged (Fig.~\ref{fig:model}~(e) and \ref{fig:model}~(f)).
Therefore, the edges of the left and right branches in Fig.~\ref{fig:model}~(b) and \ref{fig:model}~(c) is a consequence of insufficient information below the sampling resolution.
Only by accounting for more subtle configurations, one can reveal physically meaningful results, such as the occurrence of a phase transition.
%




\newpage

\bibliography{ref}




\end{document}